\def\ha{H$\alpha$}
\def\hb{H$\beta$}
\def\hd{H$\delta$}
\def\hz{H$\zeta$}
\def\he{H$\epsilon$}
\def\nt{[N\,{\sc ii}]}

\def\ntb{[N\,{\sc ii}] $\lambda\lambda 6583$}
\def\otw{[O\,{\sc ii}]}
\def\otr{[O\,{\sc iii}]}
\def\otra{[O\,{\sc iii}] $\lambda\lambda 4959$}
\def\otrb{[O\,{\sc iii}] $\lambda\lambda 5007$}

\def\nttha{[N\,{\sc ii}]/H$\alpha$}
\def\cah{Ca\,{\sc ii}-H}
\def\cak{Ca\,{\sc ii}-K}
\def\st{[S\,{\sc ii}]}
\def\sta{[S\,{\sc ii}] $\lambda\lambda 6717$}
\def\stb{[S\,{\sc ii}] $\lambda\lambda 6731$}

\def\av{$A_V$}
\def\zph{$z_{\rm phot}$}
\def\zsp{$z_{\rm spec}$}
\def\cs{$\chi^2$}

\def\figa{
  \begin{figure}[!t] 
    \begin{center}
      \includegraphics[width=0.3\textwidth]{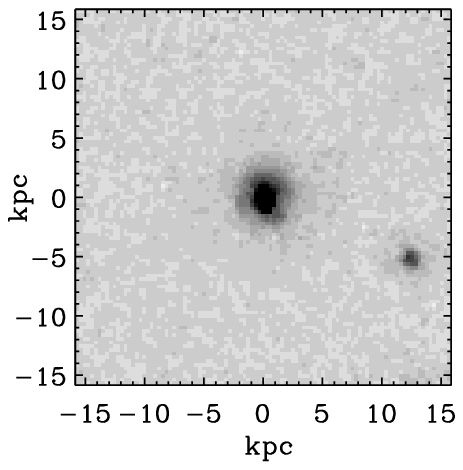}\hspace{-0.2in}
      \caption{NICMOS image of 1255-0. This image is 3\farcs8 by
        3\farcs8; thus, 1\arcsec\ corresponds to 8.3 kpc. See
        \cite{vd08b} for detailed information and for the images of a
        large sample of such compact quiescent galaxies at
        $z\sim2.3$. The galaxy is barely resolved; most of the visible
        structure is due to the point-spread function. \label{fig:im}}
    \end{center} 
  \end{figure} 
}

\def\figb{
  \begin{figure}[!t] 
    \begin{center}
    \includegraphics[width=0.48\textwidth]{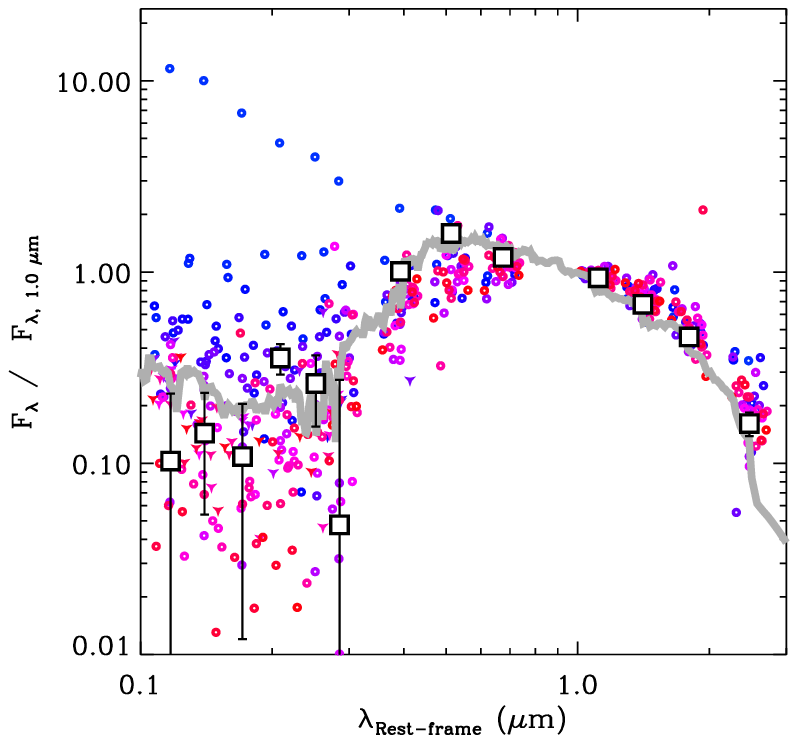}
    \caption{Comparison of the broadband photometric rest-frame UV-NIR
      SED of 1255-0 (black open squares) with the full galaxy sample
      (colored circles) from MUSYC \citep{qu07,ma09}. To match the
      redshift and stellar mass of 1255-0, we restrict the sample to
      galaxies in the range $1.9<z_{\rm phot}<2.5$ and $11.0 < {\rm
      log}(M_* / M_{\odot}) < 11.5$. The SEDs are normalized to unity
      at 1 $\rm \mu m$. The color of each SED reflects the rest-frame
      UV flux in the normalized spectrum. Triangles indicate 1$\sigma$
      upper limits. The gray curve represents the best-fit SPS model
      to the full SED of 1255-0.\label{full_sample}}
    \end{center}
  \end{figure} 
}

\def\figc{
  \begin{figure*}[!t] 
    \begin{center}
      \includegraphics[width=1.0\textwidth]{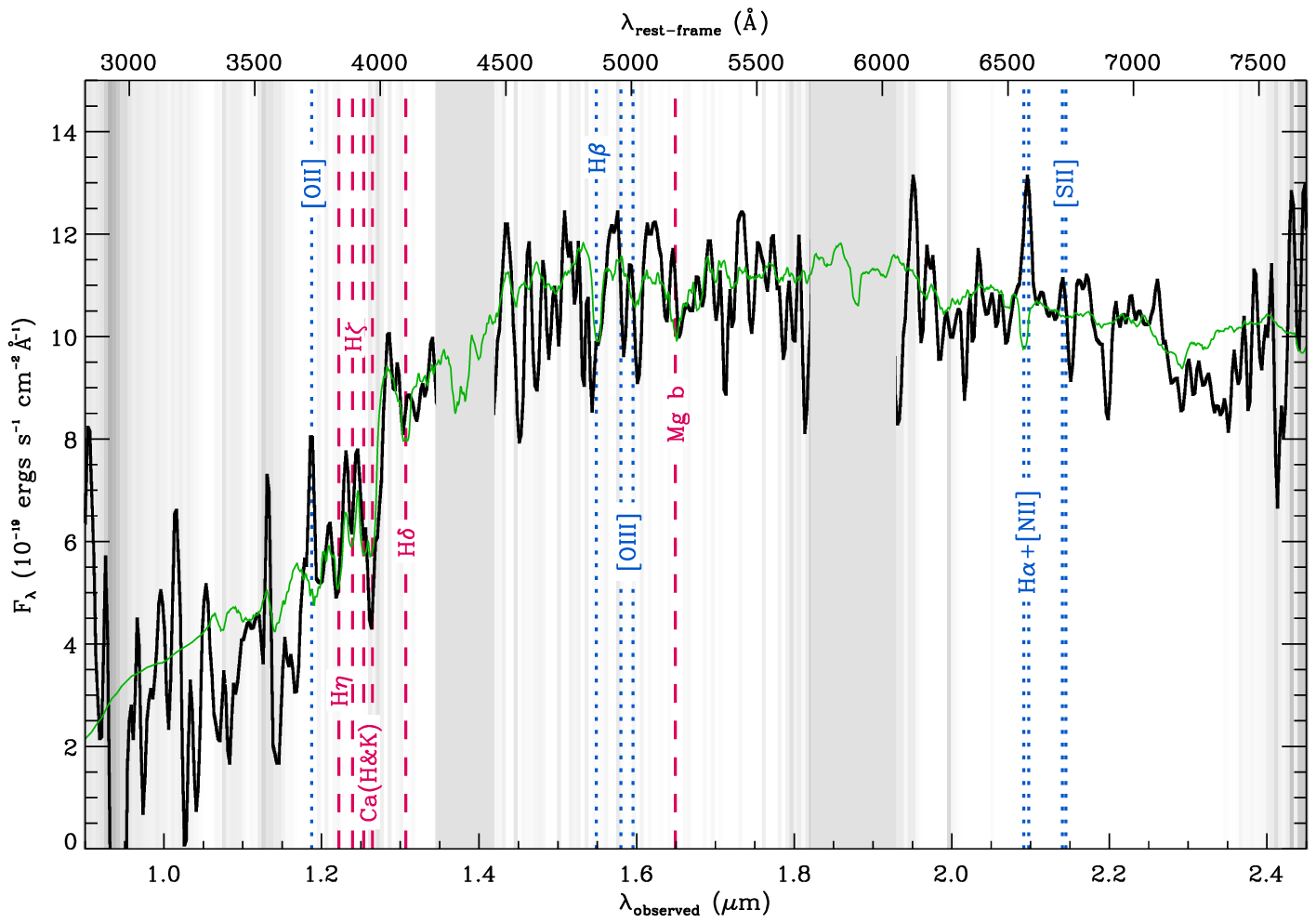}
      \caption{The $\sim29$ hr deep GNIRS spectrum of 1255-0 ({\em
          black}). The spectrum is sampled to bins of 50 \AA\ in
        observed frame, and smoothed by a boxcar of 75 \AA. The gray
        shaded background indicates the noise level of the spectrum,
        with dark being noisier. We omitted the regions in between the
        atmospheric windows. Overplotted in green is the best-fit
        \cite{bc03} SPS model to the spectrum and optical-IR
        photometry, assuming a \cite{ch03} IMF and solar
        metallicity. This fit corresponds to a stellar mass of
        $2.3\times10^{11} M_{\odot}$, an age of 2.1~Gyr, a $\tau$ of
        0.3~Gyr, a reddening of $A_v=0.25$~mag, an SFR of 1.9
        $M_{\odot}/\rm yr$, and a specific SFR of
        0.008~Gyr$^{-1}$. The locations of possible absorption and
        emission lines are indicated by the red dashed and blue dotted
        lines, respectively.  \label{full_spec}}
    \end{center}
  \end{figure*}
 }

\def\figd{
  \begin{figure}[!t] 
    \begin{center}
      \includegraphics[width=0.48\textwidth]{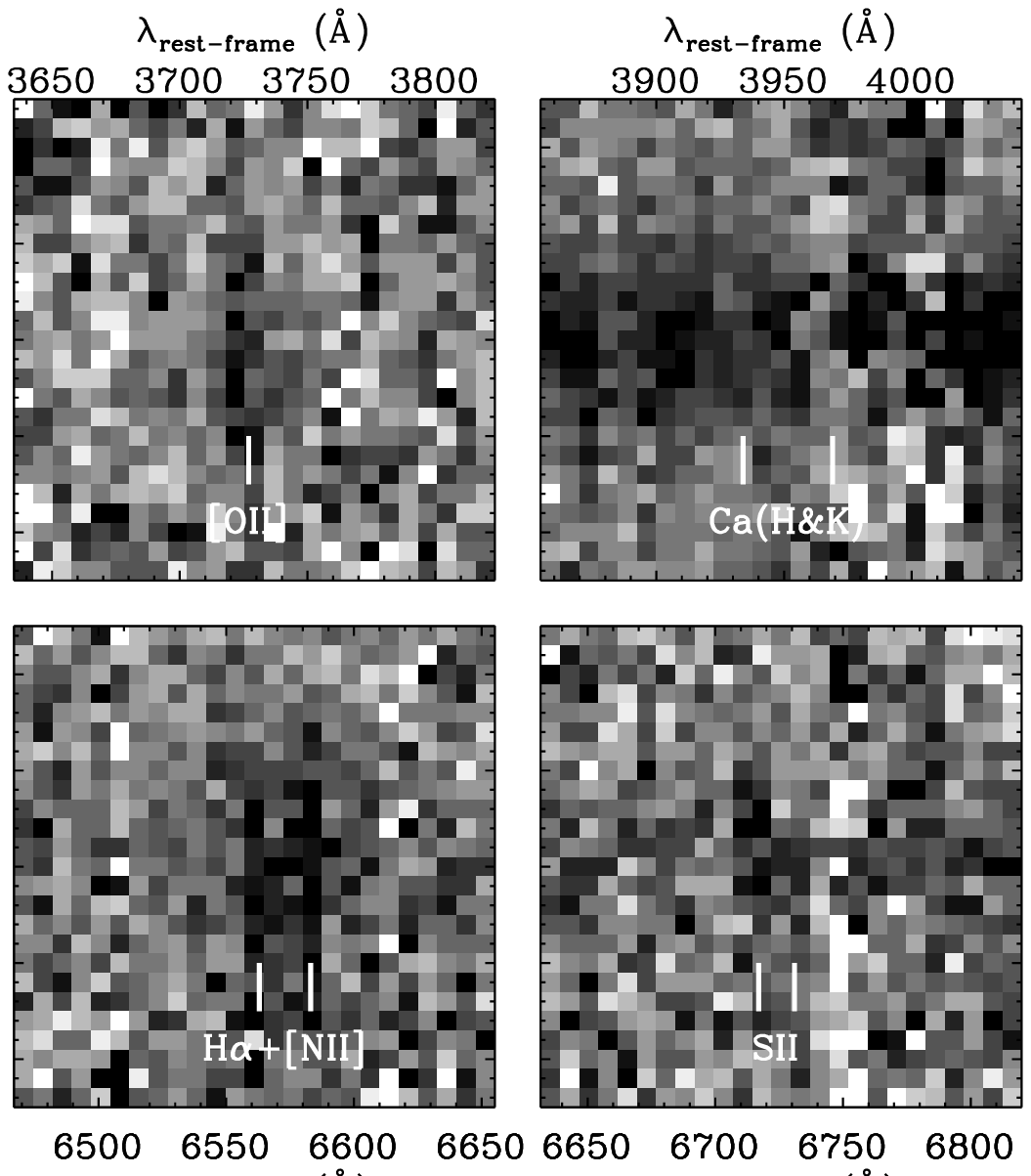}
      \hspace{-0.2in}
      \caption{Selected regions of the two-dimensional
	spectrum. Each pixel is 25\,\AA\ in observed frame (8\,\AA\
	in rest-frame) in the wavelength (horizontal) direction by
	0\farcs15 in the spatial (vertical) direction. The top-left,
	bottom-left, and bottom-right panels show the emission lines
	\otw, \ha\ and \nt, and \st, respectively. The continuum
	emission has been removed, using the best-fit model, to make
	the lines more visible. The top-right panel shows the
	wavelength region around the break. \cah\ and \cak\ are
	clearly visible in the two-dimensional spectrum. In this panel, the
	continuum emission has not been removed. \label{fig:2dlines}}
    \end{center} 
  \end{figure} 
}

\def\fige{
  \begin{figure}[!t] 
    \begin{center}
      \includegraphics[width=0.46\textwidth]{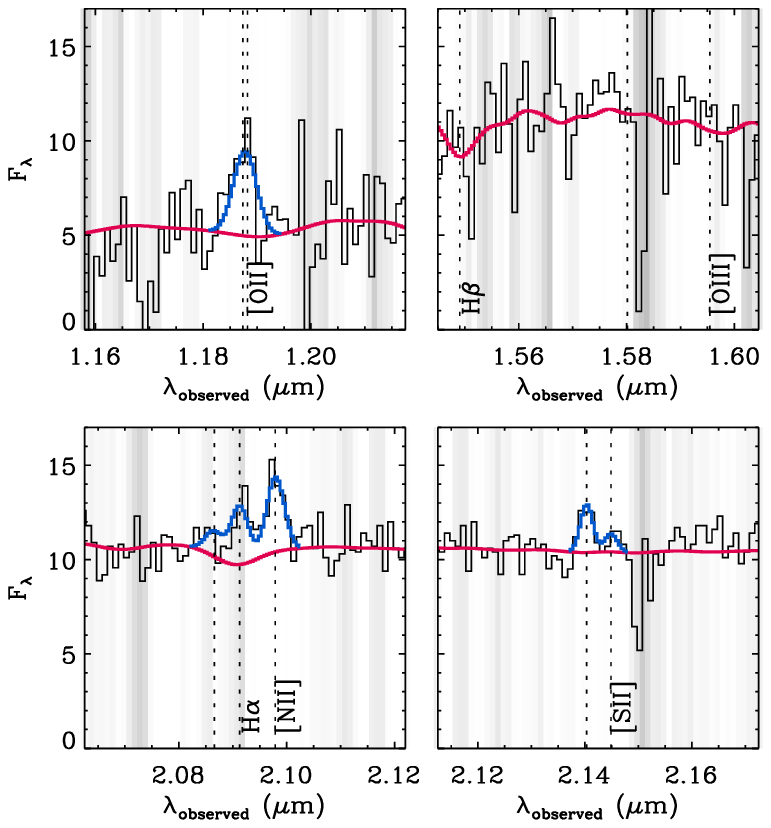}
      \caption{One-dimensional spectrum (10 \AA\ per bin in observed frame) in
        regions around the emission lines. $F_{\lambda}$ is in
        $10^{-19}$ erg\,$\rm s^{-1}\,cm^{-2}\,\AA^{-1}$. The red and
        blue lines represent the best continuum and emission-line
        fits, respectively. While we detect \otw, \ha, \nt\ and \st,
        we see no signs of \hb\ in emission or the two \otr\
        lines. \label{fig:lines}}
    \end{center} 
  \end{figure} 
}

\def\figf{
  \begin{figure}[!t] \begin{center}
    \includegraphics[width=0.5\textwidth]{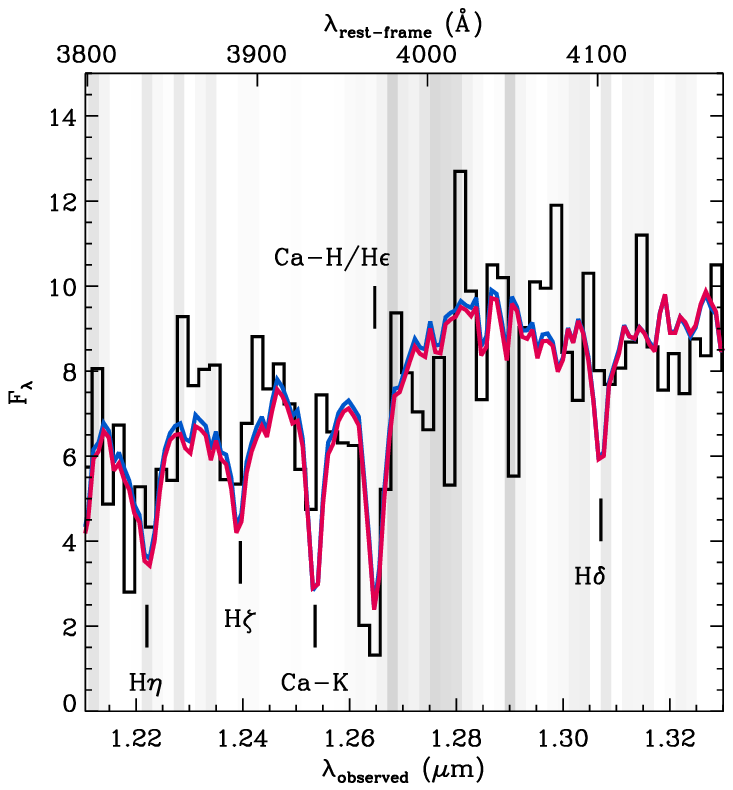}
    \caption{Spectrum (20 \AA\ per bin in observed frame) in
      the wavelength region around the rest-frame optical
      break. $F_{\lambda}$ is in $10^{-19}$ erg\,$\rm
      s^{-1}\,cm^{-2}\,\AA^{-1}$. The red and blue curves represent
      the best fits for supersolar ($2.5Z_{\odot}$) and solar
      metallicity, respectively. Both the \cah/\he\ and \cak\ lines
      are detected, as well as H$\eta$ and \hz. \label{fig:break}}
    \end{center} 
  \end{figure} 
}

\def\figg{
  \begin{figure}[!t] 
    \begin{center}
      \includegraphics[width=0.48\textwidth]{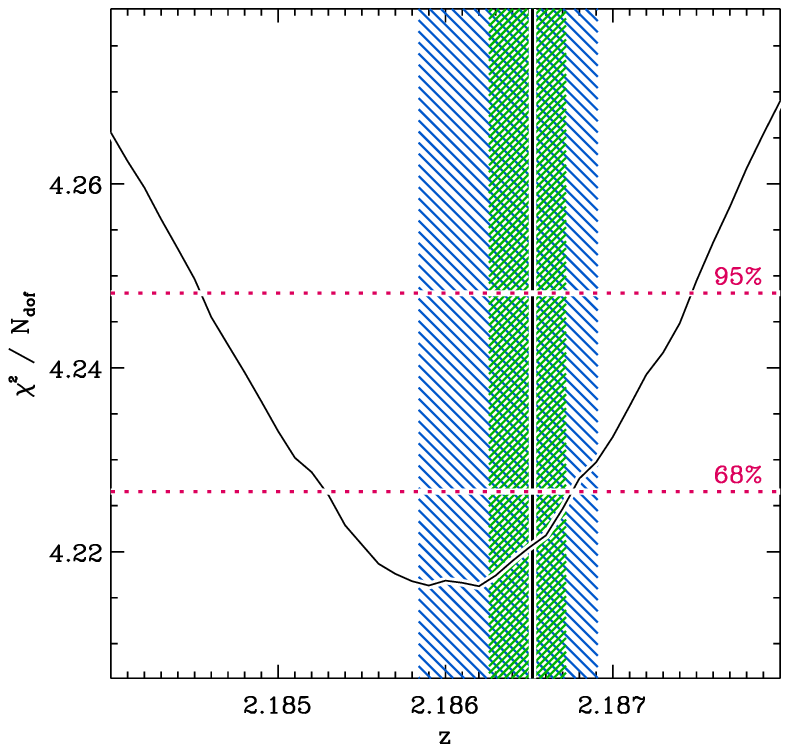}\hspace{-0.2in}
      \caption{Reduced $\chi^2$ vs. continuum redshift (number of
        degrees of freedom $N_{\rm dof}=319$), when leaving redshift
        as a free parameter in the SPS model \citep{bc03} fitting. The
        other parameters, including metallicity, are allowed to
        vary. The rest-frame UV-NIR broadband photometry is not
        included in the fit. The 68\% and 95\% confidence intervals
        are indicated by the dotted lines. The best-fit emission line
        redshift is indicated by the vertical solid line, and its 68\%
        and 95\% confidence intervals by the green and blue shaded
        regions, respectively. The continuum emission yields a well-constrained spectroscopic redshift. The redshift difference
        between the stellar and nebular emission is not at all
        significant.\label{fig:chiz}}
    \end{center} 
  \end{figure} 
}

\def\figh{
  \begin{figure}[!t] 
    \begin{center}
      \includegraphics[width=0.46\textwidth]{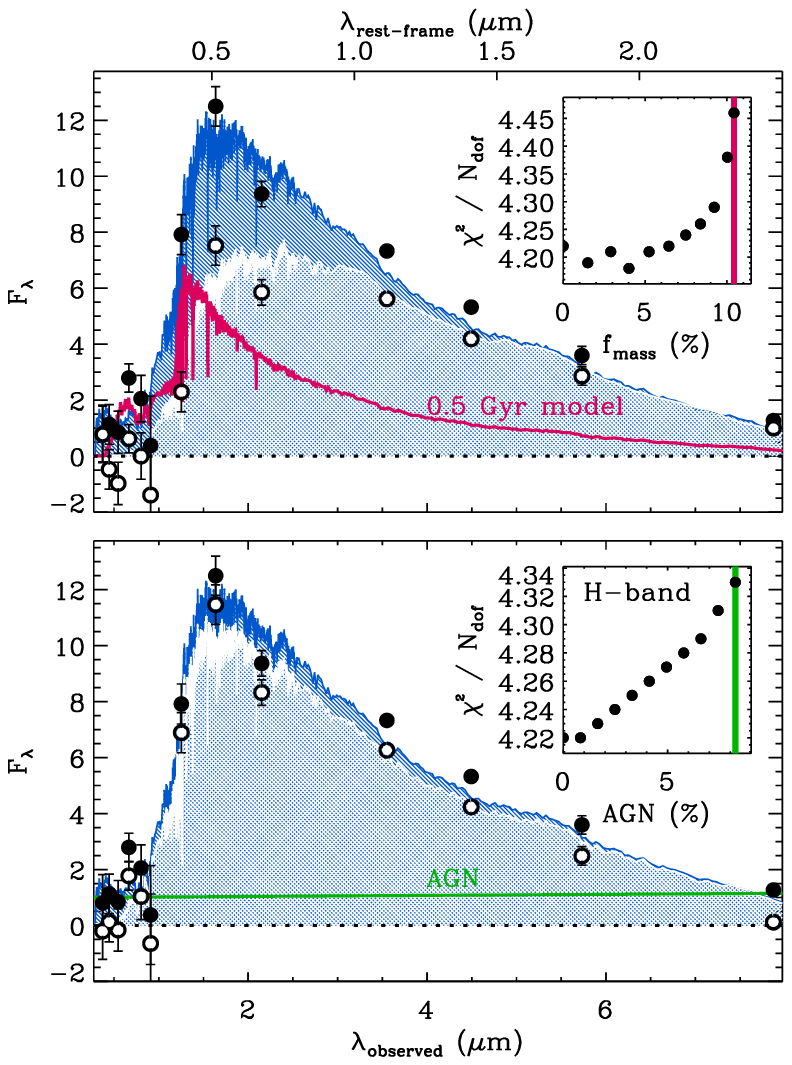}
      \caption{Multi-component modeling of 1255-0. The black solid
        dots and blue SEDs show the original broadband photometric
        data and best-fit SPS model to the full SED (including NIR
        spectrum). In the top panel we assess the contribution from a
        post-starburst population (with an \av\ of 0.25 mag), which
        can account for $<10\%$ of the stellar mass and $<40\%$ of the
        $H$-band flux. The corrected SED and corresponding best-fit
        SPS model are represented by the open symbols and white curve,
        respectively. The inset shows the quality of the fit as a
        function of the contribution of the post-starburst population
        to the total stellar mass. The red lines indicate the maximum
        contribution. In the bottom panel, we assess a contribution
        from an AGN. The inset shows the quality of the fit as a
        function of the AGN contribution to the $H$-band flux. The
        green lines indicate the maximum contribution (8\%). In both
        cases, these fits imply that the galaxy is primarily composed
        of an evolved stellar population.\label{fig:comp}}
    \end{center} 
  \end{figure} 
}

\def\taba{
  \begin{deluxetable}{l l c c}
    \tabletypesize{\scriptsize}
    \tablecaption{Observations\label{tab:obs}}
    \tablewidth{0pt} \tablehead{\colhead{} & 
      \colhead{} & \colhead{Exposure Time} & \colhead{Seeing} \\
      \colhead{ID run} & 
      \colhead{Date} & \colhead{(minutes)} & \colhead{(\arcsec)}}
    \startdata
    GS-2005A-Q-20 & 2005 May 19 & ~\,85 & 0.9 \\
                  & 2005 May 27 & ~\,80 & 1.0 \\
		  & 2005 May 30 & ~\,80 & 1.0 \\
    GS-2006A-C-6  & 2006 Feb 24 & ~\,60 & 0.5 \\
    GS-2007A-C-9  & 2007 Mar 11 & 360 & 0.5 \\ 
                  & 2007 Mar 12 & 330 & 0.4 \\ 
		  & 2007 Mar 13 & 340 & 0.5 \\ 
		  & 2007 Mar 14 & 390 & 0.6 \\
    \enddata
  \end{deluxetable}
}

\def\tabb{
  \begin{deluxetable*}{l l c c c c c}
    \tabletypesize{\scriptsize}
    \tablecaption{Spectral Features\label{tab:lines}}
    \tablewidth{0pt} 
    \tablehead{ & \colhead{$\lambda$} & \colhead{EWs} & \colhead{$F$} & 
      \colhead{$L$} & & \colhead{$\sigma$\tablenotemark{d}} \\ 
      \colhead{Lines} & \colhead{\AA} & \colhead{\AA} & 
      \colhead{ $10^{-17}$ ergs s$^{-1}$\,cm$^{-2 }$} 
      & \colhead{$10^{42}$ ergs s$^{-1}$} & \colhead{$z$} & \colhead{km\,s$^{-1}$}}
    \startdata
     \otw\tablenotemark{a} & 3727 & $14.8^{+1.8}_{-2.0}$ & 
     $2.24^{+0.31}_{-0.36}$ & $0.87^{+0.11}_{-0.13}$ & ... & 
     $522^{+88}_{-185}$\\ 
     \hb\tablenotemark{a,b,c} & 4863 & $<1.3$ & $<0.42 $ & $<0.15$ & ... & ... \\
     \otr\tablenotemark{a,b} & 4959 & $<0.4$ & $<0.14 $ & $<0.05$ & ... & ... \\
     \otr\tablenotemark{a,b} & 5007 & $<1.3$ & $<0.43 $ & $<0.15$ & ... & ... \\
     \nt & 6548 & $1.7^{+0.1}_{-0.1}$ & $0.55^{+0.05}_{-0.03}$ & 
     $0.20^{+0.02}_{-0.01}$ & $2.1865^{+0.0002}_{-0.0002}$ &
     $203^{+26}_{-15}$\\ 
     \ha\tablenotemark{c} & 6563 & $4.0^{+0.5}_{-0.6}$ & 
     $1.29^{+0.16}_{-0.18}$ & 
     $0.46^{+0.06}_{-0.06}$ & $2.1865^{+0.0002}_{-0.0002}$ & 
     $203^{+26}_{-15}$\\ 
     \nt & 6583 & $5.1^{+0.4}_{-0.3}$ & $1.66^{+0.14}_{-0.09}$ & 
     $0.60^{+0.05}_{-0.03}$ & $2.1865^{+0.0002}_{-0.0002}$ & 
     $203^{+26}_{-15}$\\ 
     \st & 6717 & $2.3^{+0.2}_{-0.3}$ & $0.76^{+0.08}_{-0.08}$ & 
     $0.27^{+0.03}_{-0.03}$ & $2.1864^{+0.0003}_{-0.0002}$ 
     & $108^{+25}_{-34}$ \\ 
     \st & 6731 & $0.9^{+0.3}_{-0.2}$ & $0.28^{+0.10}_{-0.07}$ & 
     $0.10^{+0.04}_{-0.03}$ & $2.1864^{+0.0003}_{-0.0002}$ 
     & $108^{+25}_{-34}$ \\ 
    \enddata
    \tablenotetext{a}{Fixed to best \zsp\ as derived from
    \ha\ and \nt.}
    \tablenotetext{b}{Fixed to best $\sigma$ as derived from
    \ha\ and \nt.}
    \tablenotetext{c}{Corrected for Balmer absorption.}
    \tablenotetext{d}{The resolution of GNIRS is $R\sim1000$,
      corresponding to $\sigma=128 \rm ~km\,s^{-1}$}
  \end{deluxetable*}
}

\documentclass[aaspp4,11pt,preprint,apjfonts]{emulateapj}

\shorttitle{An Ultra-Deep Spectrum of a $z=2.2$ Galaxy}
\shortauthors{Kriek et al.}

\begin{document}
  
\title{An ultra-deep near-infrared spectrum of a compact quiescent
  galaxy at $\lowercase{z}=2.2$}

\author{Mariska Kriek\altaffilmark{1}, Pieter G. van
  Dokkum\altaffilmark{2}, Ivo Labb\'e\altaffilmark{3}, Marijn
  Franx\altaffilmark{4}, Garth D. Illingworth\altaffilmark{5},
  Danilo Marchesini\altaffilmark{2}, \& Ryan F. Quadri\altaffilmark{4}}

\email{mariska@astro.princeton.edu}

\altaffiltext{1}{Department of Astrophysical Sciences, Princeton 
  University, Princeton, NJ 08544}

\altaffiltext{2}{Department of Astronomy, Yale University, New Haven, 
  CT 06520}

\altaffiltext{3}{Hubble Fellow, Carnegie Observatories, Pasadena, CA 91101}

\altaffiltext{4}{Leiden Observatory, Leiden University, NL-2300 RA Leiden, 
  Netherlands}

\altaffiltext{5}{UCO/Lick Observatory, University of California, Santa
  Cruz, CA 95064}

\begin{abstract} 

  Several recent studies have shown that about half of the massive
  galaxies at $z\sim2$ are in a quiescent phase. Moreover, these
  galaxies are commonly found to be ultra-compact with half-light
  radii of $\sim1$~kpc. We have obtained a $\sim29$~hr spectrum of a
  typical quiescent, ultra-dense galaxy at $z=2.1865$ with the Gemini
  Near-Infrared Spectrograph. The spectrum exhibits a strong optical
  break and several absorption features, which have not previously
  been detected in $z>2$ quiescent galaxies. Comparison of the
  spectral energy distribution with stellar population synthesis
  models implies a low star formation rate (SFR) of
  $1-3\rm~M_{\odot}\,yr^{-1}$, an age of $1.3-2.2$~Gyr, and a stellar
  mass of $\sim2\,\times\,10^{11}\,M_{\odot}$. We detect several faint
  emission lines, with emission-line ratios of \nt/\ha, \st/\ha,
  and \otw/\otr\ typical of low-ionization nuclear emission-line
  region. Thus, neither the stellar continuum nor the nebular emission
  implies active star formation. The current SFR is $<1\%$ of the past
  average SFR. If this galaxy is representative of compact quiescent
  galaxies beyond $z=2$, it implies that quenching of star formation
  is extremely efficient and also indicates that low luminosity active
  galactic nuclei (AGNs) could be common in these objects. Nuclear
  emission is a potential concern for the size measurement. However,
  we show that the AGN contributes $\lesssim8\%$ to the rest-frame
  optical emission. A possible post-starburst population may affect
  size measurements more strongly; although a 0.5~Gyr old stellar
  population can make up $\lesssim10\%$ of the total stellar mass, it
  could account for up to $\sim40\%$ of the optical
  light. Nevertheless, this spectrum shows that this compact galaxy is
  dominated by an evolved stellar population.

\end{abstract}


\section{INTRODUCTION}\label{sec:intro}

The first massive, quiescent galaxies ($>10^{11} M_{\odot}$) arose
when the universe was only $\sim$3 Gyr old
\citep[e.g.,][]{la05,kr06b,rw08} or perhaps even earlier
\citep[e.g.,][]{br07,man09,fo09,mo05,wi08}. Remarkably, these galaxies
already form a red sequence at $z\sim2.3$ \citep{kr08b}. The
relatively young ages ($\sim0.5$ Gyr) and post-starburst spectral
shapes \citep{kr06b,kr08b} of the $z\sim2.3$ red-sequence galaxies
suggest that a significant fraction of the stars have formed over a
short timescale in an intense starburst. Sub-millimeter bright
galaxies are possible candidates to represent this vigorous phase of
star formation
\citep[e.g.,][]{ch04,ta08,wa08}. These dusty starburst galaxies have
observed star formation rates (SFRs) of several hundreds up to a
thousand solar masses a year. The exact mechanism responsible for
transforming such active systems into quiescent galaxies is still
subject to debate
\citep[e.g.,][]{cr06,bo06,na07,de08,ho08}.

Not all local early types were already massive, quiescent systems at
these epochs \citep[e.g.,][]{vd06,ent09a}. The majority of them quench
or assemble into more massive systems at later times, and the number
density of the massive end of the red sequence at $z\sim2.3$ is only
$\sim1/8$ of the local value \citep{kr08b}. Furthermore, the future
evolution of massive quiescent galaxies at $z\sim2.3$ is still
unclear. Their evolved stellar populations suggest that they passively
evolve into their local analogs. However, their strong size and slow
color evolution contradict this picture. Recent morphological studies
show that massive quiescent systems at $z\sim2$ are remarkably compact
with effective radii of
$\sim1$~kpc \citep[e.g.,][]{tr06,tr07,zi07,to07,lo07,vd08b,ci08,fr08,dam08,we08,sa09}. Local
early types of similar stellar mass are about a factor of 5
larger \citep[e.g.,][]{vd08b}. Thus, these high-redshift galaxies must
evolve significantly after $z\sim2$, probably by inside-out growth,
primarily through minor
mergers \citep[e.g.,][]{be09,na09,we09,ho09b}. In addition, the slow
color evolution of the red sequence from $z\sim2.3$ to the present
implies that passive evolution alone cannot explain the observed
color--redshift relation \citep{kr08b}.

However, both the size and color evolution studies are hampered by
many uncertainties and detailed, crucial information on the early
phases of massive, early type galaxies is still lacking. The
constraints on the stellar populations and SFRs at high redshifts are
poor, even for galaxies with spectroscopy and mid-infrared
photometry \citep[e.g.,][]{kr08a,mu09}. Thus, it is still unclear how
``dead'' $z>2$ quiescent galaxies really are. Also, there are no
dynamical mass measurements available for quiescent galaxies beyond
$z=2$. Consequently, all stellar mass estimates are photometric, and
thus suffer from uncertainties in the derived stellar populations and
from assumptions in the metallicity and the initial mass function
(IMF). Moreover, spectroscopic redshifts are extremely difficult to
obtain for quiescent galaxies without emission lines. Using optical
spectroscopy, \cite{ci08} have derived several absorption-line
redshifts, but due to the relative faintness of quiescent galaxies in
the rest-frame UV these observations require $\sim100$ hr of
integration and result in incomplete samples. Near-infrared (NIR)
spectroscopy allows the detection of the ``bright'' rest-frame optical
continuum emission, but deep spectra are expensive due to the lack of
multiplexing, the bright NIR background, and strong OH lines. Thus, so
far no rest-frame optical absorption lines have been detected in NIR
spectra of $z>2$ quiescent galaxies, and previous redshift
determinations rely on the detection of the Balmer and/or 4000~\AA\
break with uncertainties of $\Delta z/(1+z) < 0.019$ \citep{kr08a}.

In this paper we present a $\sim29$ hr Gemini Near-Infrared
Spectrograph \citep[GNIRS:][]{el06} spectrum of a compact, quiescent
galaxy at $z=2.1865$, allowing a detailed study of the stellar
population and the detection of any rest-frame optical emission and
absorption lines. Moreover, it gives us a glance into the future, as
this is the deepest single-slit NIR spectrum ever taken of a $z>2$
galaxy. Throughout the paper, we assume a $\Lambda$CDM cosmology with
$\Omega_{\rm m}=0.3$, $\Omega_{\Lambda}=0.7$, and $H_{\rm 0}=70$~km
s$^{-1}$ Mpc$^{-1}$. All broadband magnitudes are given in the
Vega-based photometric system.

\figa 

\section{TARGET SELECTION AND DATA}\label{sec:data}

The target was chosen from our GNIRS NIR spectroscopic survey for
massive galaxies at $z\sim2.3$ \citep{kr08a}. All galaxies were
originally selected from the Multi-Wavelength Survey by Yale-Chile
(MUSYC), which provides us with deep optical-IR photometry
\citep{ga06,qu07,ent09b,da09,ma09}. These ``shallow'' spectra
\citep[typically $\sim$3 hrs of integration, see][]{kr08a} allowed us
to derive continuum redshifts and classify the galaxies.

We selected 1255-0 from the nine massive quiescent galaxies presented
in \cite{kr06b}, because of its redshift (the optical break falls in
the $J$ band), and its visibility at the time of our GNIRS run. The
galaxy is not brighter than the other candidates. Also its effective
radius ($r_e$) of 0.78 kpc is very similar to the median $r_e$ (0.9
kpc) of the other massive, quiescent galaxies in our sample
\citep[see][]{vd08b}. Figure~\ref{fig:im} shows the image of
1255-0 as obtained by the Near-Infrared Camera and Multi-Object
Spectrometer (NICMOS). Altogether, this galaxy seems typical for the
general class of quiescent galaxies at $z\sim2.3$.

\figb 

We also consider how this galaxy compares to the general population of
massive galaxies at similar redshift (including star-forming
galaxies). In Figure~\ref{full_sample} we compare the rest-frame
UV-NIR spectral energy distributions (SEDs) of all galaxies in the
deep MUSYC sample \citep{qu07,ma09} with comparable redshift and
stellar mass. The redshifts are all photometric, and derived using
EAZY \citep{br08}. The stellar masses are derived using the code
described in the Appendix, for the \cite{ma05} stellar population
synthesis (SPS) models, solar metallicity, the \cite{ca00} reddening
law, and a \cite{kr01} IMF. Figure~\ref{full_sample} illustrates that
1255-0 is slightly redder than the average massive galaxy at this
redshift. This is expected given the fact that it was selected to be
quiescent.

\taba
\figc 

In total we integrated nearly 29 hr on 1255-0 with GNIRS (with
individual exposures of 10 minutes) divided over three observing runs
in 2005 May, 2006 February, and 2007 March. The integration times and
average seeing are given in Table~\ref{tab:obs}. During the first
three nights the conditions were mediocre, with occasional clouds and
an average seeing of $1\arcsec$. The weather conditions were excellent
during the last two runs, with clear skies and an average seeing of
0.5$\arcsec$.

The galaxy was observed in a cross-dispersed mode, in combination with
the 32 lines~mm$^{-1}$ grating and the 0\farcs675 slit. The spectral
resolution varies between $\sim$900 and $\sim$1050 over the different
orders. Observations were done using an ABA\arcmin B\arcmin\ on-source
dither pattern, such that we can use the average of the previous and
following exposures as sky frame \citep{vd04,kr06a}. Acquisition was
done using blind offsets from nearby stars. Before and after every
observing sequence we observe an A\,V0 star, for the purpose of
correcting for telluric absorption. The final spectra of the two stars
are combined to match the airmass of the observing sequence.

A detailed description of the reduction procedure of GNIRS
cross-dispersed spectra is given in \cite{kr06a}. In summary, we
subtract the sky, mask cosmic rays and bad pixels, straighten the
spectra, combine the individual exposures, stitch the orders, and
finally correct for the response function. The different observing
sequences are weighted according to their signal-to-noise ratio (S/N)
when being combined.

A one-dimensional spectrum is extracted by summing all adjacent lines
(along the spatial direction) with a mean flux greater than 0.1 times
the flux in the central row, using optimal weighting with the S/N. We
extract both a high- and low-resolution spectrum with 10 \AA\ and
50 \AA\ per bin, respectively. The high-resolution spectrum, which is
resampled such that no resolution is lost, is used for spectral
features, while the low-resolution spectrum is extracted to study the
continuum emission. In order to flux calibrate the spectrum, we derive
the spectroscopic $J$, $H$, and $K$ fluxes by integrating over the
corresponding filter curves. We derive one scaling factor by comparing
the spectroscopic colors with the $J$, $H$, and $K$ broadband
photometric data from MUSYC. Finally, we multiply the spectrum by this
scaling factor.

The low-resolution spectrum is shown in Figure \ref{full_spec}. Note
the clear detection of the rest-frame optical break in the $J$ band
and the relatively high S/N in the continuum \citep[compared to the
spectrum shown in][]{kr06b}.

\section{ANALYSIS}

We study the properties of this galaxy in two ways, first by measuring
and analyzing the spectral features (Sections \ref{sec:abs} and
\ref{sec:ion}) and second by modeling and decomposing the stellar
continuum emission (Sections \ref{sec:con}--\ref{sec:agn}). Finally,
in Section \ref{sec:comp}, we compare the deep spectrum modeling
results with our previously published shallow spectrum.

\subsection{Emission and Absorption Features}\label{sec:abs}\label{sec:em}

\tabb 
\figd 
\fige 

We measure the redshift and all emission-line properties by modeling
the extracted high-resolution (10\,\AA\ per bin) one-dimensional
spectrum. We detect \ha, \ntb, and \sta\ in the $K$ band. The bottom
panels of Figures~\ref{fig:2dlines} and \ref{fig:lines} show the
relevant part of the two-dimensional and one-dimensional spectrum,
respectively. For the former we first removed the best-fit continuum
model (see Section \ref{sec:con}) to make the lines more visible. We
detected none of these lines in our shallow spectrum \citep{kr06b}, as
they are too faint.

We model \ha\ and the two \nt\ lines simultaneously, by assuming the
same redshift and width for all three lines, the best-fit continuum
model as derived in Section \ref{sec:con}, and Gaussian profiles. As the
detected emission lines are all faint, it is important to accurately
correct for continuum emission (including the Balmer absorption
lines). The continuum model is corrected for the spectral resolution
of GNIRS and convolved to the same velocity width as the emission-line
model. Furthermore, we adopt the ratios of transition probability
between the two \nt\ lines of 0.34.

We derive the best values for the redshift, line width, and the fluxes
of the emission lines by minimizing \cs. The uncertainties on the
modeling results are derived using 500 Monte Carlo simulations. In the
simulations, we perturb the measured spectrum using the noise
spectrum. The results are listed in Table~\ref{tab:lines}. \sta\ and
\stb\ are modeled in the same way; thus the redshift, width, and
scaling for both lines were free parameters.

In the $H$ band we detect no emission lines (see Figure \ref{fig:lines}
top-right panel). We derive 2~$\sigma$ upper limits using 500 Monte
Carlo simulations in which we model the lines assuming the width and
redshift as obtained from the \ha\ and \nt\ lines, and the continuum
model. We fit all three expected lines \hb, \otra\ and \otra\
simultaneously, assuming a ratio of 0.33 between the two \otr\
lines. Upper limits are derived from the best-fit results of all
simulations.

In the $J$ band we detect the doublet \otw\ (see
Figures \ref{fig:2dlines} and \ref{fig:lines}, top-left panels). We do
not resolve the two lines separately, and thus we fix the ratio (to 1)
and redshift when fitting the lines. The combined fitting parameters
are given in Table~\ref{tab:lines}.

For each pair of lines we fit simultaneously, we assume that all line
emission originates from the same physical region. \st\ and \ha\
yielded consistent redshifts, but the different combinations of lines
resulted in different widths. This might imply that the lines
originate from different processes and regions in the galaxy. However,
the lines are all faint, and the errors on the line widths are
probably somewhat underestimated because of our fitting procedure
(allowing more freedom in the fits would result in larger
uncertainties, particularly in the linewidths). Furthermore, we do not
know the relative contributions from the two \otw\ line. Thus, we
cannot draw any firm conclusion from the different line widths.

\figf 

As far as we are aware of, this is the first rest-frame optical
spectrum of a quiescent galaxy beyond $z=2$ for which rest-frame
optical absorption lines are detected. Figure~\ref{fig:break} shows
that we detect H$\eta$, \hz, \cak\ and \cah/\he. \cak\ and \cah/\he\
are also visible in the top-right panel of
Figure~\ref{fig:2dlines}. In Figure~\ref{fig:break}, we also show the
best-fit \cite{bc03} models for fixed solar metallicity and when
leaving the metallicity as a free parameter
(see Section~\ref{sec:con}). Furthermore, the absorption lines Mg b
$\lambda 5175$ and \hb\ are detected in the low-resolution spectrum in
Figure~\ref{full_spec}. There might also be a hint for \hd\ in
Figure~\ref{full_spec}. However, a significant detection of this line
requires an even deeper spectrum.

This deep spectrum allows an accurate measure of the strength of the
4000~\AA\ break, which is an indicator of the evolution stage or age
of the stellar population. We use the definition of \cite{ba99} to
measure the strength of this break, and find a value of
$D_n(4000)=1.40^{+0.03}_{-0.03}$. The corresponding age depends on the
star formation history (SFH), the metallicity, and the dust content of
the galaxy. For the extreme case of a simple stellar population (SSP)
model with solar metallicity, an age of $\sim0.6$~Gyr is needed to
produce such a break. Assuming little or no dust, this value can be
seen as a lower limit on the population age. A more detailed
measurement of the age of the stellar population will be given in
Section~\ref{sec:con}.

  \begin{deluxetable*}{l c c c c c c c} \tabletypesize{\scriptsize}
    \tablecaption{Stellar population properties
      \label{tab:mod}} \tablewidth{0pt} \tablehead{ & \colhead{Log
        $M_*$} & \colhead{Log age} & \colhead{Log $\tau$} &
      \colhead{\av} & &
      \colhead{Log SFR } & \colhead{Log SFR / $M_*$} \\
      \colhead{$UBVRIz$ + NIR spectrum} & \colhead{($M_{\odot}$)} &
      \colhead{(yr)} & \colhead{(yr)} & \colhead{(mag)} & \colhead{$Z$} &
      \colhead{($M_{\odot} \rm \, yr^{-1}$)} & \colhead{(Gyr$^{-1}$)}}
    \startdata              BC03, Chabrier, $Z_{\odot}$       & $   11.36^{+    0.00}_{-    0.04    }$ & $    9.34^{+    0.02}_{-    0.04    }$ & $     8.5^{+     0.0}_{-     0.1    }$ & $    0.20^{+    0.15}_{-    0.05    }$ & $    0.02^{+    0.00}_{-    0.00    }$ & $    0.19^{+    0.16}_{-    0.20    }$ & $   -2.17^{+    0.17}_{-    0.18     }$ \\
                BC03, Chabrier, $Z$ free       & $   11.26^{+    0.00}_{-    0.02    }$ & $    9.10^{+    0.02}_{-    0.02    }$ & $     8.3^{+     0.0}_{-     0.1    }$ & $    0.30^{+    0.05}_{-    0.05    }$ & $    0.05^{+    0.00}_{-    0.00    }$ & $    0.45^{+    0.00}_{-    0.36    }$ & $   -1.81^{+    0.00}_{-    0.34     }$ \\
     BC03, Chabrier,  $Z_{\odot}$, +IRAC       & $   11.36^{+    0.00}_{-    0.04    }$ & $    9.32^{+    0.04}_{-    0.04    }$ & $     8.5^{+     0.0}_{-     0.1    }$ & $    0.25^{+    0.10}_{-    0.10    }$ & $    0.02^{+    0.00}_{-    0.00    }$ & $    0.28^{+    0.07}_{-    0.29    }$ & $   -2.08^{+    0.08}_{-    0.27     }$ \\
          BC03, Chabrier, $Z$ free +IRAC       & $   11.26^{+    0.00}_{-    0.02    }$ & $    9.10^{+    0.02}_{-    0.02    }$ & $     8.3^{+     0.0}_{-     0.1    }$ & $    0.30^{+    0.00}_{-    0.10    }$ & $    0.05^{+    0.00}_{-    0.00    }$ & $    0.45^{+    0.00}_{-    0.36    }$ & $   -1.81^{+    0.00}_{-    0.34     }$ \\
               MA05, Kroupa, $Z_{\odot}$       & $   11.24^{+    0.00}_{-    0.02    }$ & $    9.26^{+    0.00}_{-    0.02    }$ & $     8.4^{+     0.0}_{-     0.1    }$ & $    0.00^{+    0.05}_{-    0.00    }$ & $    0.02^{+    0.00}_{-    0.00    }$ & $    0.01^{+    0.11}_{-    0.42    }$ & $   -2.23^{+    0.12}_{-    0.40     }$ \\
         MA05, Kroupa, $Z_{\odot}$ +IRAC       & $   11.24^{+    0.00}_{-    0.04    }$ & $    9.26^{+    0.00}_{-    0.04    }$ & $     8.4^{+     0.0}_{-     0.1    }$ & $    0.00^{+    0.10}_{-    0.00    }$ & $    0.02^{+    0.00}_{-    0.00    }$ & $    0.01^{+    0.20}_{-    0.42    }$ & $   -2.23^{+    0.21}_{-    0.40     }$ \\
 \enddata \tablecomments{When deriving
      the confidence intervals we only use the model grid points
      enclosed by the 68\% $\chi^2$ threshold values (see
      Appendix). Due to this discretization, non-zero confidence
      intervals may be rounded to zero.}
  \end{deluxetable*} 

\subsection{Ionization Mechanism}\label{sec:ion}

Emission-line ratios can be used to study the origin of the ionized
emission. In particular, \nttha, which reflects both the metallicity
and ionization parameter $U$ of a galaxy, is a powerful
discriminator. Ratios of \nttha~$>$~1 suggest that an object is
ionized by a hard radiation field; H\,{\sc ii} regions are not able to
produce such high ratios \citep{ke01}. The ratio of
\st/\ha\footnote{For \st\ we take the sum of both lines, while for the
  ratio of \nttha\ only \ntb\ is used} can be used for a similar
purpose, as H\,{\sc ii} regions have \st/\ha~$<0.6$. The high values
for both \nttha\ and \st/\ha\ of $1.3^{+0.3}_{-0.2}$ and
$0.8^{+0.2}_{-0.2}$ respectively show that normal star-forming regions
are not the dominant contributor to the line emission in this galaxy.

The ratio of \otr/\otw\ can be used to further characterize the
hardness of the ionizing radiation field \citep{sh90,kd02}. \otrb\ is
not detected in our spectrum, but we derive a 2\,$\sigma$ upper limit
on this parameter. We find log(\otr/\otw) $<-0.89$, implying an
ionization parameter $U<10^7$ \citep{ke01}. This ratio is consistent
with the original definition for low-ionization nuclear emission-line
regions (LINERs) of \otw/\otr $>$ 1, and so the line emission in
1255-0 is most likely caused by a LINER.

The origin of LINER emission has been a subject of debate since its
original classification as an active galactic nucleus (AGN) class
by \cite{he80}. Although there is substantial evidence that many
LINERs are powered by accretion onto massive black holes, LINER
emission can also originate from a young starburst or by shock heating
through cloud collisions induced by galaxy mergers or starburst-driven
winds
\citep[e.g.,][]{ds95}. Because the SFR in 1255-0 is very low (see
Section~\ref{sec:con}), a starburst or starburst-driven wind is
unlikely to cause the LINER emission, and thus a low-luminosity AGN is
the more likely option. Nonetheless, even if the observed LINER
emission does not originate from black hole accretion, it is still the
case that normal star-forming regions do not dominate the observed
line emission. 

Finally, we note that the line widths are not necessarily indicative
of the depth of the gravitational potential, as in the local universe
there is a large scatter between the gas line widths of LINERs and the
velocity dispersion of the stars of their host galaxies
\citep{gr05}.

\subsection{Modeling the Continuum Emission}\label{sec:con}

We study the nature of the stellar population by comparing the
spectrum with the SPS models of \cite{bc03} and \cite{ma05}, using the
code described in the Appendix. We assume an exponentially declining
SFH with timescale $\tau$, a \cite{ch03} or \cite{kr01} IMF, and
the \cite{ca00} reddening law. We derive model spectra for a grid with
$\tau$ between 10 Myr and 10 Gyr in steps of 0.1 dex, \av\ between 0
and 3 mag in steps of 0.05 mag, and age in steps of 0.02 dex with a
minimum age of 10 Myr and the maximum age not exceeding the age of the
universe. The metallicity when fitting the \cite{ma05} model is fixed
to solar ($Z=0.02$), but for the \cite{bc03} models we vary the
metallicity, among subsolar ($Z=0.004$), solar, and supersolar
($Z=0.05$). The redshift is fixed to $z=2.1865$ as derived from the
emission lines (see Section~\ref{sec:em}). We fit the low-resolution
one-dimensional spectrum (50 \AA\ per bin in observed frame) and
search for the best solution by minimizing
\cs.

In contrast to our previous studies \citep{kr06a,kr06b,kr08a}, we do
not mask regions with low atmospheric transmission or strong sky
emission when fitting the spectrum. As the S/N of this spectrum is
considerably higher than in \cite{kr08a}, we apply a much smaller bin
size. Bins in bad wavelength regions will have larger uncertainties,
and thus simply have lower weight in the fit. When using larger bins
\citep[400 \AA\ in][]{kr08a} this method is less appropriate, as a
bad region will contaminate nearly all bins.

To further constrain the SFR, we include the rest-frame UV broadband
photometry. Furthermore, we extend the SED into the rest-frame NIR
using IRAC photometry \citep{ma09}. The rest-frame NIR helps to
constrain the stellar mass and age of the galaxy
\citep[e.g.,][]{la05,sh05,wu07,mu09}. However, as the SPS models are
still uncertain in this regime, we fit the galaxy both with and
without the IRAC photometry.

We derive 68\% confidence intervals on all stellar population
properties using 200 Monte Carlo simulations, as described in the
Appendix. The photometric uncertainties are increased using the
template error function, which accounts for uncertainties in the model
templates as a function of the rest-frame wavelength
\citep{br08}. Furthermore, we apply the automatic scale option, such
that for each simulation the simulated spectrum is calibrated using
the simulated $J$, $H$, and $K$ photometry (see Section~\ref{sec:data}
and the Appendix). Thus, the uncertainty in the calibration of the
spectrum is explicitly taken into account. 

In Table~\ref{tab:mod} we give all modeling results for the different
SPS libraries, free or fixed metallicity, and with or without IRAC. In
Figure~\ref{full_spec}, we show the spectrum and best-fit \cite{bc03}
model with solar metallicity. The continuum fitting implies a stellar
mass of $\sim2\times10^{11}~M_{\odot}$, a reddening of
$A_V=0.0-0.3$~mag, a star formation timescale of $\tau\sim0.3$~Gyr, an
age (since the onset of star formation) of $1.3-2.2$~Gyr, and an SFR
of $1-3 ~M_{\odot}\,\rm yr^{-1}$. The universe is 3 Gyr at
$z\sim2.1865$, which implies a formation redshift $z_{\rm form}=4-7$.

It is remarkable how well constrained the formal confidence intervals
are. However, the formal errors do not reflect the true uncertainties
properly, as they are dominated by the systematic effects such as the
assumptions concerning the SPS models, the SFH, metallicity, and
extinction law \citep[see e.g.,][for more discussion on this
topic]{sh01,wu07,kg07,co08,mu09}. For the \cite{bc03} models,
supersolar metallicity is formally preferred above solar. We do not
believe the fact that this result is significant, due the strong
degeneracy between age and metallicity, and the uncertainties in the
SPS model, and so the metallicity could well be less, e.g.,
solar. Nonetheless, the different models are surprisingly
consistent. The inclusion or exclusion of IRAC has little effect on
the modeling results for this deep spectrum
\citep[see][for a general discussion of the inclusion of IRAC data in
the modeling of $z\sim2.3$ galaxies with shallower
spectroscopy]{mu09}.

\figg 

The detection of the strong break and absorption features allows the
measurement of a stellar continuum redshift. We use the same fitting
procedure as described above; only this time, we leave the redshift as a
free parameter. Furthermore, we fit only the spectrum; thus, the
rest-frame UV and NIR broadband photometric data are not included. In
Figure~\ref{fig:chiz}, we show the reduced $\chi^2$ as a function of
redshift. We find a stellar redshift of
$z=2.1862^{+0.0005}_{-0.0009}$. The emission-line redshift is also
indicated in this figure. There is no evidence for a significant
redshift offset between the stellar and nebular emission. Nonetheless,
we cannot exclude a possible offset, due to the relatively large
uncertainties on the continuum redshift.

\figh 

\subsection{A Recent Starburst?}\label{sec:bur}

When modeling the SED by SPS models, we assume an exponentially
declining SFH. However, this is a simplification, and more complex
SFHs are more realistic. For example, subsequent merging is expected
to result in central dissipational starbursts
\citep[e.g.,][]{ho09}. This would result in a younger stellar
population in the central part of the galaxy, with a lower
mass-to-light ratio ($M/L$) than the older underlying
population. Similarly, if the galaxy experienced recent star formation
in a disk-like component, the galaxy is composed of an old central
concentration and a more extended young component. This younger
population will contribute relatively more to the observed light than
to the stellar mass.

In order to assess whether a recent starburst took place and how much
it may contribute to the stellar mass of the galaxy, we investigate
how much of the light can be accounted for by a 0.5 Gyr SSP model,
with 0.25 mag of visual extinction (see Table~\ref{tab:mod}). We
choose a post-starburst instead of an ongoing starburst, as we know
that the current global SFR in the galaxy is very low. We subtract
different mass contributions and apply the same fitting procedure as
discussed in Section~\ref{sec:con}. The maximum contribution of the
post-starburst population is set by the rest-frame UV flux, and shown
by the red SED in the top panel of Figure~\ref{fig:comp}. Up to
$\sim$10\% of the stellar mass may have been formed in a recent
starburst. Due to the relatively low $M/L$ of this post-starburst
population, the contribution to the $H$-band light is much larger
($\sim40\%$).

In the inset of the top panel of Figure~\ref{fig:comp} we show the
reduced $\chi^2$ of the best-fit SPS model to the corrected spectrum
and photometry, for different mass fractions of the post-starburst
stellar population. Models with a mass fraction between $\sim0$ and
$5\%$ provide equally good fits, while higher mass fractions result in
worse agreement. The red line indicates the maximum contribution,
corresponding to the red SED. Thus, while this galaxy may have
experienced a recent starburst, both the light and the stellar mass
are dominated by an older stellar population. Nonetheless, it still
remains to be explored how much the light distribution may be
different from the stellar mass distribution in these compact
quiescent galaxies.

\subsection{Continuum Emission from an AGN?}\label{sec:agn}

In Section \ref{sec:ion}, we found that the line emission of 1255-0 is
of LINER origin. This raises the question of whether an AGN may
contribute to the continuum emission of 1255-0. We investigate this by
subtracting different AGN contributions and fitting the corrected SED
by SPS models. We assume a power-law SED for the AGN. The maximum AGN
contribution is set by the rest-frame UV fluxes in combination with
the 8 $\mu$m IRAC band, and shown by the green line in the bottom
panel of Figure~\ref{fig:comp}. The corresponding contribution to the
$H$-band flux is $\sim8\%$.

In the inset of the bottom panel of Figure~\ref{fig:comp} we show the
reduced $\chi^2$ of the best-fit SPS model for different assumed AGN
contributions. The green line indicates the maximum contribution,
corresponding to the green SED. The fit clearly worsens for an
increasing AGN contribution. Altogether, an AGN is unlikely to
contribute significantly to the continuum emission, and is limited to
a maximum of $\sim$\,8\% to the $H$ band.

\subsection{Comparison to the shallow spectrum}\label{sec:comp}

While our 5 hr\footnote{The weather conditions during the first runs
  were significantly worse, and the effective exposure time is closer
  to 1-2 hr.} shallow spectrum of 1255-0 presented in \cite{kr08b}
  provided a similar stellar mass and SFR (when corrected for the
  difference in the assumed IMF), the age, the dust content, and the
  continuum redshift differ by $\sim2\,\sigma$ compared to the 29 hrs
  deep spectrum. Our shallow spectrum of 1255-0 yields a continuum
  redshift of $2.31^{+0.05}_{-0.07}$, an age of $0.57^{+0.44}_{-0.28}$
  Gyr, and an \av\ of $1.2^{+0.6}_{-0.6}$ mag. Thus, this galaxy was
  previously classified as a dusty post-starburst galaxy, at slightly
  higher redshift. In the fit to the shallow spectrum the dominant
  optical break was thought to be the Balmer jump, while in our deeper
  spectrum the 4000~\AA\ break is found to be the more prominent
  one. If we fit the shallow spectrum of 1255-0 with the redshift
  fixed to $z=2.1865$, we find a significantly older age of 2.9 Gyr
  and a lower
\av\ of 0.35 mag.

The change in redshift also influences the rest-frame color
determination \citep[see][]{kr08b}. 1255-0 was among our reddest
galaxies, with a rest-frame $U-B$ color of $0.36^{+0.05}_{-0.06}$
mag. The deep spectrum yields an $U-B$ color of $0.52$ mag. However,
if we use the same method as in \cite{kr08b}, thus determining the
color of the best fit, we find $U-B=0.30$. This difference is likely
caused by the discrepancy between the spectrum and the fit around 1.15
$\mu$m.

We do not expect the full sample of shallow spectra to suffer as
severely from this degeneracy between redshift and stellar population
as 1255-0. In \cite{kr08a} we apply the same continuum fitting
procedure to the emission-line galaxies in our sample, and find a good
agreement between the emission-line and continuum redshifts. This
sample contained several galaxies with SED shapes similar to those
without emission lines. Nonetheless, while it seems more likely that
1255-0 is the largest outlier ($\sim2\sigma$), caution is
required. This case illustrates that more deeper spectra are needed.

\section{THE STAR FORMATION ACTIVITY IN 1255-0}

Comparison of the stellar continuum emission with SPS models
(Section~\ref{sec:con}) confirms our earlier results that the SFR in
1255-0 is strongly suppressed \citep{kr06b}. Depending on the SPS
library and the assumed metallicity, the best-fit SFR is
$1-3~M_{\odot}\, \rm yr^{-1}$. As expected for a galaxy with low-level
star formation, the total dust extinction is low with values of \av\
of 0.0-0.3 mag.

The low SFR is independently confirmed by the emission-line
diagnostics. In the previous section, we noted that star-forming
regions cannot be the dominant contributor to the line
emission. Nevertheless, if we assume that the detected \ha\ emission
is caused by just star formation, we can use the calibration
\begin{equation}
  \label{eq:kennicutt} {\rm SFR} \ (M_{\odot}\,{\rm yr^{-1}}) =
  7.9 \times 10^{-42} \ L_{\rm H\alpha} \ {\rm (erg\,s^{-1})}
\end{equation}
as given by \cite{ke98} to derive the SFR from the \ha\
luminosity. Assuming no dust extinction, the observed luminosity would
result in an SFR of $4.0^{+0.4}_{-0.6}\,M_{\odot} \rm \, yr^{-1}$ for
a \cite{sa55} IMF. This value would decrease by a factor of
$\sim$\,1.8 for a \cite{ch03} or \cite{kr01} IMF, but increases by a
factor of $\sim2.3$ if there is 1 mag of visual extinction in the star
forming regions. Although we have a good constraint on the total
attenuation of the galaxy from the spectral modeling
(\av~$=0.0-0.3$~mag), it is difficult to estimate the extinction in
the line-emitting regions without measuring a Balmer decrement. The
combination of \ha\ and the 2\,$\sigma$ upper limit on \hb\ gives a
2\,$\sigma$ lower limit on the ratio of \ha/\hb\ of 3.07. This limit
is close to the intrinsic ratio for H\,{\sc ii} regions of 2.76, and
thus sets no constraints on the dust content. Nonetheless, even when
assuming that the line emission originates just from star formation,
and assuming 1 mag of dust extinction in the H\,{\sc ii} regions, the
obtained SFR is still only $\sim4~M_{\odot}~\rm yr^{-1}$ (for a
Chabrier IMF).

The best-fit stellar mass of 1255-0 is $1.7-2.3 \times 10^{11}
M_{\odot}$. This value is slightly dependent on the choice for the SPS
model, and on whether metallicity is assumed to be solar or left as a
free parameter. If left as a free parameter, supersolar is preferred
over solar, although we argue in Section~\ref{sec:con} that this result is
not significant. The best-fit star formation timescale is
$0.2-0.3$~Gyr, and the best-fit stellar age is $1.8-2.2$~Gyr and
$\sim1.3$~Gyr for solar metallicity and supersolar metallicity,
respectively. The age of the universe at this redshift is 3.0 Gyr and
the formation redshift is $z_{\rm form}\sim4-7$.

We study the evolution stage of the galaxy by means of the quenching
factor $q_{\rm sf}$, which measures by what factor the SFR has been
reduced compared to the past average SFR.  We define $q_{\rm sf}$ as
follows:
\begin{equation}
  \label{eq:q} {q_{\rm sf} = \rm 1 - \frac{SFR_{current}}{SFR_{past}}
    = 1 - \frac{SFR_{current}}{\frac{\it M_*}{age}}}
\end{equation}
In this equation $M_*$ is not the current stellar mass, but all the
mass formed in the galaxy. This quenching factor is related to the
ratio of the age to the star formation timescale $\tau$ (i.e., how
many e-folding times have been passed since the galaxy was formed) in
the following way
\begin{equation}
  \label{eq:q} {q_{\rm sf} =  1
    - \frac{age}{\tau} \frac{e^{-\frac{age}{\tau}}}{1 -
    e^{-\frac{age}{\tau}}}}
\end{equation}
Because age/$\tau$ is a relatively well-constrained parameter, $q_{\rm
  sf}$ is a fairly robust measure for the evolution stage of a
  galaxy. We note that \cite{da08} defined a comparable factor: the
  star formation activity parameter $\alpha_{\rm sf} = (M_{*}/{\rm
  SFR}) / (t_H-1 {\rm Gyr)}$. Instead of using the best-fit age of the
  galaxy, this parameter uses the Hubble time ($t_H$) minus 1 Gyr.

1255-0 has a $q_{\rm sf}$ of 0.991$^{+0.007}_{-0.002}$. Thus, the SFR
in this galaxy has been reduced by more than 99\% and the current SFR
is less than 1\% of the average past SFR. This implies that the star
formation has been strongly quenched since the major star formation
epoch. Furthermore, in Section~\ref{sec:bur} we found that a 500~Myr
SSP accounts for $<$10\% of the stellar mass, which implies that at
least 90\% of the stellar mass is in stars older than 0.5 Gyr, and
thus formed beyond $z=2.6$.

\section{DISCUSSION AND CONCLUSIONS}

Due to the introduction of large photometric surveys with deep NIR imaging
\citep[e.g.,][]{la03,fr03}, many quiescent massive galaxies beyond
$z=2$ have been identified in the past few years
\citep[e.g.,][]{fo04,da04a,da04b,la05}. Moreover, these galaxies are
typically found to be ultra-compact, with stellar densities that are
about 2 orders of magnitude larger than in local early type galaxies
of similar mass \citep[e.g.,][]{tr06,tr07,zi07,to07,lo07,vd08b,ci08}

Follow-up spectroscopic studies have tried to verify the broadband
photometric redshifts and stellar population properties of these
ultra-dense quiescent galaxies. This turned out to be extremely
difficult, and with optical spectroscopy tens to hundreds of hours are
required \citep{da05,ci08}, due to their relative faintness in the
rest-frame UV. Evolved galaxies beyond $z=2$ are much brighter at NIR
wavelengths, corresponding to the rest-frame optical. In \cite{kr06b},
we used the Balmer and / or 4000~\AA\ breaks to obtain redshift
estimates for a sample of nine massive quiescent galaxies. However, exact
redshift measurement from rest-frame optical absorption lines remained
out of reach until this paper.

By integrating for nearly 30 hr with GNIRS we succeeded in detecting
for the first time rest-frame optical absorption lines in an NIR
spectrum of a compact, quiescent galaxy beyond $z=2$. This deep
spectrum has full NIR coverage ($\sim$\,1.0-2.4 $\mu$m). In addition to
the absorption features H$\eta$, \hz, \cak, \cah/\he, \hb, and Mg~b,
we have detected \otw, \ha, \nt, and \st\ in emission. All emission
lines are faint with luminosities of
$0.1--0.9\times10^{42}\rm\,ergs\,s^{-1}$. The redshifts derived from
the stellar continuum and from the emission lines are consistent
within the uncertainties.

Comparison of the spectral continuum emission and the rest-frame
UV-NIR photometry with SPS models implies a stellar mass of
$\sim2\times10^{11} M_{\odot}$, a reddening of \av$=0.0-0.3$~mag, a
star formation timescale of $\tau=0.2-0.3$~Gyr, and an age of
$1.3-2.2$~Gyr. The results are slightly different for the different
SPS models and different assumed metallicities. We find a low SFR of
about $1-3~M_{\odot}\,\rm yr^{-1}$, implying that the star formation
is strongly quenched and reduced by more than 99\% since its major
star formation epoch. If this galaxy is typical for quiescent
galaxies, quenching of star formation is extremely efficient. The
constraints on the SFR are very tight, and compared to the previously
published shallow spectrum of this galaxy \citep[][]{kr08a} they are
more stringent by a factor of $\sim$8.

We do detect a faint \ha\ emission line, which, based on emission-line
diagnostics, is not caused by stellar ionization. However, even if we
assume that the \ha\ line was due to star formation, it would result
in a low SFR of $\sim2-4~M_{\odot}\,\rm yr^{-1}$.  One possibility is
that obscured star formation may have been missed. However, at low
redshift there is a strong correlation between the dust-corrected
luminosity of \ha\ and the bolometric luminosity
\citep[e.g.,][]{mo06}. \cite{re06} found a similar relation at high
redshift. Thus, based on the faint \ha\ line, the nature of the line
emission, and the lack of a detection at 24\,$\mu$m (I. Cury, private
communication), we do not expect this galaxy to host obscured ongoing
starburst regions.

Although we can confirm our earlier conclusions about the quiescent
nature of the stellar population in 1255-0 \citep{kr06b}, we
previously underestimated the age. The shallower spectrum showed an
apparent Balmer break and our fit preferred a younger dusty
post-starburst galaxy. However, our deeper spectrum shows that the
galaxy is $1.3-2.2$~Gyr old, and the optical break is dominated by the
4000\,\AA\ break rather than the Balmer jump. A post-starburst
(0.5~Gyr) population can only account for $\lesssim10$\% of the
stellar mass.

The underestimated age in the shallow spectrum is related to the
overestimation of the continuum redshift. In \cite{kr08a} we found a
continuum redshift of $2.31^{+0.05}_{-0.07}$, which is within
2$\sigma$ consistent with the emission-line redshift of 1255-0. This
may imply that the uncertainties in our previous work are
underestimated. However, comparison of emission-line and continuum
redshifts for 19 emission-line galaxies in \cite{kr08a} demonstrated
the reliability of the uncertainties of our continuum redshifts, and
thus 1255-0 is probably ``the'' largest (2\,$\sigma$) outlier.

The fact that we previously underestimated the age of the galaxy shows
how difficult it is to estimate ages based on ``shallow'' spectra, let
alone broadband photometry. Thus, we cannot exclude that the ages of
more quiescent galaxies may have been systematically underestimated in
\cite{kr08a}. The best-fit models imply a formation redshift of
$z=4-7$ for 1255-0, and this may indicate that massive galaxies with
strongly suppressed star formation exist at even earlier times.

The rest-frame optical emission-line diagnostics indicate that 1255-0
most likely hosts a LINER. In \cite{kr07} we found that at least 4 out
of the 11 emission-line galaxies in our massive galaxy sample at
$2.0<z<2.7$ host an AGN. This study was based on relatively shallow
spectroscopy ($\sim3$ hr per galaxy). For 1255-0, we did not detect
any emission lines in the shallow NIR spectrum \citep{kr06b}. Thus, we
may find actively accreting black holes in more massive galaxies, when
we obtain deeper data. This suggests the possibility that low
luminosity AGNs may be very common in these objects.

Although an AGN dominates the line emission, its contribution to the
continuum emission is very low. In the $H$ band, in which we measured
the compact size of 1255-0 \citep{vd08b} the contribution is
$\lesssim8$\%. Thus, an AGN could not be the dominant cause for the
compact size of this galaxy. A central post-starburst population may
have a larger effect on our size measurements. If the stellar
population in the center has a lower $M/L$ than the outskirts, the
size will appear smaller. Similarly, if the galaxy experienced recent
star formation in the outer parts, the $M/L$ will be higher in the
center, and thus the size will appear larger.  Although a
post-starburst population can only account for $\lesssim10$\% of the
stellar mass, it can make up $\sim40\%$ of the $H$-band flux. Thus,
depending on where this post-starburst population is situated, the
light distribution in these compact galaxies may be more or less
concentrated than the stellar mass distribution, and our size
measurements may be over- or underestimated. However, regardless of
whether or not a post-starburst or AGN is present, this spectrum shows
that this compact quiescent galaxy is not actively forming new stars,
and is primarily composed of an evolved stellar population.

The future generation of NIR spectrographs promises higher throughput,
in combination with multiplexing. This will allow even deeper spectra
of larger samples. However, the most spectacular results -- in
particular with regard to kinematic measurements of quiescent galaxies
-- are expected to come from NIRSPEC on the {\it James Webb Space
Telescope (JWST)}. Without the hindrance of the Earth's atmosphere, it
will be possible to obtain significantly deeper spectra in only a
small fraction of the exposure time needed for the spectrum presented
in this paper.

\acknowledgments We thank the members of the MUSYC collaboration for
their contribution to this work, I. Cury and A. Muzzin for their help
with the MIPS imaging, G. Brammer for the photometric redshifts of the
MUSYC sample, and J. Greene and A. van der Wel for helpful
discussions. This work is based on observations obtained at the Gemini
Observatory, which is operated by the Association of Universities for
Research in Astronomy, Inc., under a cooperative agreement with the
NSF on behalf of the Gemini partnership, and on observations made with
the {\it Spitzer Space Telescope}, which is operated by the Jet Propulsion
Laboratory, California Institute of Technology under a contract with
NASA. Support from NASA grant HST-GO-10808.01-A is gratefully
acknowledged. G.D.I. acknowledges support from NASA grant
NAG5-7697. D.M. is supported by NASA LTSA NNG04GE12G. R.F.Q. is
supported by a NOVA postdoctoral fellowship.

\appendix
\section{Fitting and Assessment of Synthetic Templates}

We developed a custom IDL code named FAST, to fit SPS models to
broadband photometry, spectra, or both. FAST is compatible with the
photometric redshift code EAZY \citep{br08}, such that the format of
the input photometric catalog and filter files is similar. Optionally,
the photometric redshifts as derived by EAZY can be read in and used
by FAST.

Summarized, FAST reads in a parameter file, which defines the
photometric and spectroscopic catalogs, the SPS models \citep[choice
from][]{bc03,ma05}, the IMF\citep[choice from][]{sa55,kr01,ch03}, the
reddening law, and the fitted grid of stellar population
properties. For all properties (age, star formation timescale $\tau$,
dust content \av, metallicity and redshift) the minimum and maximum
value, and the step size can be defined. Optionally, FAST can
calibrate spectra using the broadband fluxes.

FAST generates a six-dimensional cube of model fluxes for the full
stellar population grid and all filters, spectral elements, or
both. To determine the best-fit parameters, it simply determines the
\cs\ of every point of the model cube. In case spectroscopic or
photometric redshifts are provided, the redshift will be fixed to the
closest value in the grid.

The confidence levels are calibrated using Monte Carlo
simulations. The number of simulations can be defined in the parameter
file. The observed fluxes are modified according to their photometric
and spectroscopic errors in each simulation. Optionally, a rest-frame
template error function \citep[see EAZY, ][]{br08} can be added to the
broadband photometric fluxes, to account for uncertainties in the
models. In case the automatic scale option is used, the spectrum is
scaled individually for each simulation. By doing this, we incorporate
the error on the scaling factor\footnote{Note that we did not apply
  this method in \cite{kr08a}; instead, we arbitrarily increased the
  uncertainties by quadratically adding 10\% of the average flux of
  the spectrum to account for systematic effects.}. We determine the
best solution for all simulations and define the $\chi^2_{1\sigma}$
($\chi^2_{2\sigma}$ or $\chi^2_{3\sigma}$) level as the \cs\ value in
the originally grid that encloses 68\% (95\% or 99\%) of the
simulations. The uncertainties on the stellar population properties
are the minimum and maximum values that are allowed within the
$\chi^2_{1\sigma}$ ($\chi^2_{2\sigma}$ or $\chi^2_{3\sigma}$) threshold.

In case photometric redshifts (as provided by EAZY) are assumed, the
calculation of the confidence intervals is slightly more
complicated. Both codes use different template sets and thus produce
different probability distribution functions of $z$. Unfortunately,
there is no perfect method to determine the confidence levels for this
case. We use the following method, which incorporates the confidence
levels of EAZY, and is fast and user friendly. We run all Monte Carlo
simulations at the best-fit \zph\ as derived by EAZY. We determine
$\Delta\chi^2$; the difference between the \cs\ value that encloses
68\% (or 95\% or 99\%) of the simulations and the minimum \cs\ value
at this \zph. Next, we return to the full \cs\ grid that resulted from
FAST. The confidence intervals for a given parameter are the minimum
and maximum values in the full model cube with a \cs\ value less than
$\chi^2_{\rm min}+\Delta\chi^2$ {\em and} a redshift within the 68\%
confidence interval as provided by EAZY. In essence, we use the EAZY
output to limit the redshift range of the solutions allowed in
FAST. Note that $\chi^2_{\rm min}$ of the full grid is likely lower
than the minimum value of \cs\ at \zph. We tested this method using a
large number of Monte Carlo simulations that are input in both EAZY
and FAST (retaining the combinations of
\zph\ and perturbed fluxes). In this case we do not use the
simulations for calibration, but derive the uncertainties directly
from the output. We find good correlations between the uncertainties
of the two methods and we believe that the FAST confidence levels are
robust and reliable.



\end{document}